\documentclass[onecolumn,showpacs,preprintnumbers,amsmath,amssymb]{revtex4}
\usepackage{graphicx}
\usepackage{dcolumn}
\usepackage{bm}
\usepackage{subfigure}
\usepackage{caption}
\usepackage{ulem}

\begin{document}
\title{Efficient priority queueing routing strategy on mobile networks\\}
\author{Ganhua Wu$^{1,2}$}
\author{Jiahui Pan$^2$}
\author{Huijie Yang$^1$}
\address{$^1$ Business School, University of Shanghai for Science and Technology,Shanghai 200093,China\\
$^2$ School of Software, South China Normal University, Guangzhou 510641, China}

\begin{abstract}
Mobile networks are intriguing in recent years due to their practical implications. Previous routing strategies for improving transport efficiency have little attention what order should the packets be forwarded, just simply used first-in-first-out queue discipline. Here we apply priority queueing discipline to shortest distance routing strategy on mobile networks. Numerical experiments show it not only remarkably improves  network throughput and the packet arriving rate, but also reduces average end-to-end delay and the rate of queueing delay. Our work may be helpful in routing strategy designing on mobile networks.
\end{abstract}



\maketitle

\section{Introduction}
\label{intro}
Traffic is widespread in engineering and social systems\cite{ChenSurvey}. For example, information delivery over the World Wide Web\cite{SatorrasPRL} and the transportation on highway system\cite{Milton}. The rapid development of society led to a huge increase of traffic volume in many real networked communication systems\cite{Ohira}. An essential challenge is how to enhance transport efficiency. It's found that the optimal performance on networks depends mainly on the structural characteristics and the routing strategy\cite{Boguna}. But for the existing networks, the structure has been fixed, such as the Internet, even if its structure is far from optimal, it is difficult to change. Hence, more effort to improve transport efficiency are in routing strategies.

The shortest path routing strategy is widely used in real communication systems\cite{SP}. In shortest path strategy, a packet walks to its destination along the shortest path. It will easily lead to congestion on hub nodes as most packets tend to pass through the links with high betweenness\cite{Danila}.
This fact consequently motivates researchers have proposed many other improved versions, which are considered global or local information on networks. For example, Yan et al. proposed efficient routing strategy that each node have the whole network¡¯s global topological information\cite{Yan2006}. They found that network capability is improved more than 10 times by optimizing the efficient path. However, the strategy based on global information may be practical for small or medium size networks but not for large real communication such as the Internet, World Wide Web due to large storage capacity in each node and heavy communication cost on searching global information in networks\cite{Albert}.

Therefore, scientists proposed many other strategies based on local information. Wang et al. proposed local information routing strategy, in which packets are routed only based on local topological information according neighbor's degree of each node with a tunable parameter $\beta$. They found maximal capacity corresponds to $\beta=-1$ in the case of identical nodes¡¯ delivering ability\cite{Wang_Local}. Liu et al. proposed an adaptive local routing strategy based on real-time load information, in which the node adjusts the forwarding probability with the dynamical traffic load(packet queue length) and the degree distribution of neighbouring nodes. They found it can improve the transmission capacity by reducing routing hops\cite{Liu_CPB}. Wang et al. proposed  mixing routing strategy by integrating the degree of node and the number of packets with parameter. It's found this strategy can improve the efficiency comparing with the strategy by adopting exclusive local static information\cite{Wang_Mix}.

With the development of mobile technology, networks of mobile agents are widely used. A typical example is mobile ad-hoc network, in which agetns move randomly and two agents can transfer data packets with each other only when the distance between them is less than a critical value\cite{Abolhasan,Camp}.
Yang et al. proposed a random routing strategy on mobile networks, in which packets deliver to a randomly selected agent in the communication circle. They found an algebraic power law between the throughput and the communication range with an exponent determined by the speed\cite{YangHX2011}. Moreover, Yang and Tang proposed an adaptive routing strategy, in which incorporates geographical distance with local traffic information through a tunable parameter. They found there exists an optimal value of the parameter, leading to the maximum traffic throughput of the network\cite{yanghx2014}.

In previous work, queue discipline has little attention in routing strategies. Queue discipline refers to the manner in which packets are selected when arrival packets exceed maximal processing capacity of agent\cite{Gross}. The most common discipline is first-in-first-out. However, this is certainly not the only possible queue discipline. The second is last-in-first-out discipline. A typical example is the last clicked information may be served first in the World Wide Web. The third is priority schemes, in which packets are given priorities, the ones with higher priorities to be selected ahead of those with lower priorities, regardless of their time of arrival to the agent. For example, paid packets with high priority are delivered first when downloaded from a certain web site.

In most existing work, first-in-first-out discipline is widely used. It is simple and convenient, but far from optimal. AS our knowledge, only a few articles use other discipline. Tadi\'c et al. study the web-graph model with last-in-first-out discipline\cite{Tadic1, Tadic2, Tadic3,PRE2015}.
Kim et al. introduce a priority routing strategy, in which each packet is pre-set a priority. They found the traffic behavior is improved in the congestion region, but worsened in the free flow region\cite{KimPriQue}. Tang and Zhou define an effective distance by considering simultaneously the waiting time and remaining path-length to the destination, according to which the packets queue in a descending order of the effective distance. They found it can remarkably enhance the network throughput\cite{Tang2011}. Du et al. propose a shortest-remaining-path-first queueing strategy in which a packet's priority is determined by the distance between its current location and destination. They found the traffic efficiency is greatly improved, especially in the congestion state\cite{Du2013}. Zhang et al. introduce a dynamic-information-based queueing strategy, They found the network capacity has no obvious change, but it got significant improvement for some traffic indexes such as average end-to-end delay, rate of queueing delay\cite{ZhangDIB}.

In previous studies, few researchers have applied priority discipline to routing strategy on mobile networks. However, in the real dynamic network, packets are often assign priority. For example, important military information often has a higher delivered priority on mobile ad hoc networks in the battlefield environment. In this article, we propose a shortest-distance-first(SDF) routing strategy on mobile networks, in which packet is delivered first if the distance between location of neighbor agent and its destination is shortest. Compared with first-in-first-out queueing discipline, our strategy not only remarkably improves network throughout and the packet arriving rate, but also reduces average end-to-end delay and the rate of queueing delay.

\section{Methods and Materials}
\subsection{Network model and queueing strategy}
Traffic is simulated on a random network of mobile agents, i.e., $N$ agents (numbered from $1$ to $N$) move on a square-shaped cell of size $L\times L$. Periodic boundary condition is used. Initially, agents distribute randomly in the area. At each time step $\Delta t$, moving direction of an agent is re-directed randomly, while its speed $v$ is selected to be a constant for simplicity. At the same time, a total of $R$ packets are generated in the network, whose source and destination are randomly selected. All the agents have a same communication radius $\alpha$. Two agents can realize a communicate with each other only when the distance between them is less than $\alpha$.

The queueing strategies are described as follows:

(1) Parameters setting. We set the number of agents $N=800$, delivery capability of each node $C=1$, time step $\Delta t=1$, total simulation time $T=5000$, and the queue buffer of each agent is unlimited.

(2) At each time, $R$ packets with random sources and destinations are generated in the network.

(3) Update the mobile agents position as follows:
\begin{equation}
\begin{split}
&x_i(t) = x_i(t-1) + \upsilon \cos \theta _i(t-1),\\
&y_i(t) = y_i(t-1) + \upsilon \sin \theta _i(t-1).\\
&i=1,2,\cdots,N.
\end{split}
\end{equation}

where $\theta_i(t-1)$ is the angle of moving direction of the $i$th agent at time $t-1$ with respect to x-axis, generated by sampling from a uniform distribution in $[-\pi, \pi]$, $x_i(t)$ and $y_i(t)$ are the coordinates of $i$th agent at time $t$.

(4) For each agent, take out $M$ packets in the queue. Let the number of packets in the queue is $K$.

    \quad(a) shortest-distance-first strategy: M = K.

    \quad(b) first-in-first-out strategy: if $C>K$, then $M=K$, otherwise, $M=C$.

(5) For each agent, check first $M$ packets in normal order. If packet's destination is within it's communication radius, the packet is sent directly to the destination and removed immediately from the system, and set $C=C-1$. If $C=0$, go to step (2) for next T.

(6) For a packet $m$ in the system, $D_m(t)$ is defined as the shortest distance between location of neighbor agent and destination of the packet, namely,

\begin{equation}
D_m(t) = \min\limits_{k \in K }  \sqrt{[x_k(t)- x_l(t)]^2 + [y_k(t) - y_l(t)]^2 },
\end{equation}

where $m=1,2, ... M$ represents the packet of agent $i$, $K$ is the set of neighbors of agent $i$, $x_k(t)$ and $y_k(t)$ represent the coordinates of the neighbor agent $k$ at time $t$, $x_l(t) $and $y_l(t)$ are the coordinates of the packet's destination at time $t$.

(7) If there are no more than C packets in the queue, all the packets are sent to their neighbor agents according to $D_m(t)$. otherwise,

    \quad(a) Shortest-distance-first strategy: The first C packets are delivered according to $D_m(t)$ in ascending order.

    \quad(b) first-in-first-out strategy: The first C packets are delivered in normal order.

(8) Repeat steps (4)--(7) for each agent.

(9) Repeat steps (2)--(8) T times.

(10) Compute the order parameter $\eta(R)$, the critical packet generating rate $R_c$ and the average end-to-end delay $T$, etc.

\subsection{Critical packet generating rate}
In the present studies the traffic is determined by two competitive factors. The one is the removing packets determined by routing strategy, the communication radius, density of agents, moving speed, and the number of packets transferred. The other is the number of packets produced each time step. When the generating rate of packets is small, new packets can arrive quickly their destination, the load will keep unchanged or even zero, called a free-flow state. When the rate increases to a certain value, averagely at each time step there appear some new packets that can not be delivered to their destination on time. This aggregation of new packets will increases rapidly the load of the network. In reality, a network has a limited capacity, a persistent overload on which will lead to onset of a congestion, i.e., a collapse of the system.
To characterize the throughput of a network, we exploit the order parameter $\eta$ introduced in Ref\cite{Arenas},

\begin{equation}
\begin{split}
&\eta (R) \equiv \lim\limits_{ t \to \infty } { { C\over R} { {\langle N_p(t+\Delta t)-N_p(t) \rangle} \over {\triangle t} }}
= \lim\limits_{ t \to \infty } { C\over R} {\langle N_p(t+1)-N_p(t) \rangle},
\end{split}
\end{equation}

where $N_p(t)$ represents the total number of packets existing in the whole network at time $t$.  When $R$ is less than a critical value of $R_c$, there is a balance between the generated and removed packets, which implies $\eta(R)=0$. When $R$ becomes larger than $R_c$, a transition occurs from a free-flow state to a congestion. A higher $R_c$ corresponds to a better algorithm.

\subsection{Average end-to-end delay}
End-to-end delay refers to  the time taken for a packet to be transmitted across a network from source to destination. It is defined as

\begin{equation}
D_{end-end} = K*(D_{trans} + D_{proc} + D_{prop} + D_{queue}),
\end{equation}

where $K$ is the number of links, $D_{trans}$ represents time to send bits into link, $D_{proc}$ represents nodal processing time, $D_{prop}$ represents propagation delay, and $D_{queue}$ represents time waiting in the queue. Here we focus on $D_{prop}$ and $D_{queue}$, thus neglect $D_{trans}$ and $D_{proc}$ for simplicity. Average end-to-end delay is defined as
\begin{equation}
<D> = (\sum_{i=1}^{N_{arrive}}{D_i})/N_{arrive},
\end{equation}

where $D_{i}$ represents end-to-end delay of packet $i$, $N_{arrive}$ is the number of arrived packets. Average end-to-end delay $<D>$ is an important measurement of a network's performance. In real communication networks, packets have a finite life time to avoid wasting the network resources\cite{Chen2010,Du2011}. For example, an error of packet's destination may cause the packet to be transmitted endlessly. Therefore, if a packet has been transferred  more than finite life time, it will be removed from the networks even if it has not achieved its destination. Obviously, high value of $<D>$ means more packets will be removed before they can achieve their destination.

\subsection{The rate of queueing delay}
The rate of queueing delay is defined as
\begin{equation}
Q = (\sum_{i=1}^{N_{arrive}}{D_{i-queue}/D_{i-end-end}})/N_{arrive},
\end{equation}

where $D_{i-end-end}$ is the end-to-end delay of packet i, $D_{i-queue}$ is the total delay in the queues of packet i, and $N_{arrive}$ is the number of arrived packets. In general, $Q$ reflects the degree of customer satisfaction. In many systems, such as airlines systems and the World Wide Web, users become impatient if $Q$ is large.

\subsection{Packet arriving rate}
Packet arriving rate reflects to the number of generated packets divided by the number of arrived packets. It is defined as
\begin{equation}
A = N_{arrive}/N_{create},
\end{equation}

where $N_{arrive}$ is the number of arrived packets and $N_{create}$  the number of generated packets. Obviously, $A$ is an index of system throughput. In the free-flowing state, the generated packets are sent to the destination on time, so $A$ is quite close to 1, while in the congestion state, the generated packets accumulate in the network and can not be distributed timely, thus $A$ is smaller than 1.

\section{Results}
Figure 1 shows two typical results of $\eta$ versus $R$ with a selection of $\alpha=1$ at low ($v=0.1$) and high ($v=1$) moving speed, respectively. For each specific case, there exists a finite value of $R_c$, at which a transition from free-flow to congestion occurs in a sharp interval of $R$. For description convenience, we represent the rapid increase of $\eta$ by using $R_c$ at which $\eta$ starts to be non-zero. We can find the $R_c$ of the shortest-distance-first strategy is larger than that of the first-in-first-out strategy, especially at high moving speed. In fact, $R_c$ of the first-in-first-out strategy is $\sim30$ at $v=1$, while that of the shortest-distance-first strategy is $\sim700$ . Obviously, the shortest-distance-first strategy remarkably improves the network throughput.

\begin{center}
\begin{figure}
\centerline{\includegraphics[width=12cm]{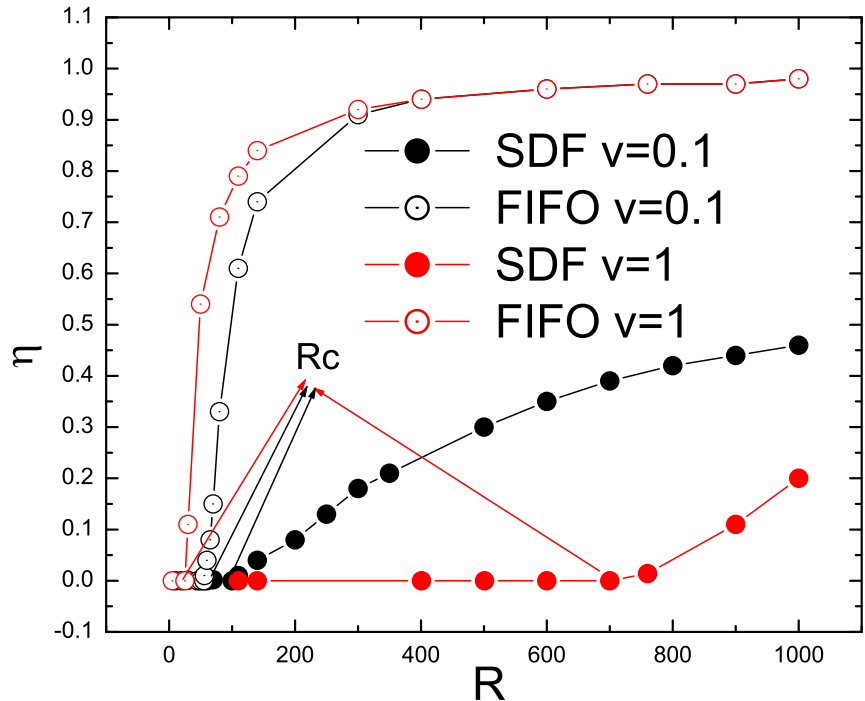}}
\begin{flushleft}
\caption{Order parameter $\eta$ versus $R$ for different values of speed $v$. The number of agents is $800$, the size of the square region $L\times L=10\times 10$, and the communication radius $\alpha=1$. Delivery capacity of each agent is $C=1$.  Each data point results from an average over $20$ realizations. Error bar is not shown as the values of error are very small.}
\end{flushleft}

\end{figure}
\end{center}
Figure 2(a) shows the dependence of $R_c$ on $v$ at $\alpha=1$ and Fig. 2(b) shows the relationship of $R_c$  and $\alpha$ at $v=0.3$. We can find  that $R_c$ of the shortest-distance-first strategy is larger than that of the first-in-first-out strategy at different speed $v$ and communication radius $\alpha$. From Fig. 2(a),  $R_c$ of the shortest-distance-first strategy increases rapidly with $v$ increases  in the beginning, and keeps stable ($\sim 700$) when $v$ is large than $0.7$, while that of the  first-in-first-out strategy increases with $v$ increases and reaches the maximum at $v = 0.1$ and then decreases slowly. Figure 2(b) represents $R_c$ of two strategies increases with the communication radius $\alpha$ increases, but $R_c$ of the shortest-distance-first increases faster than that of the first-in-first-out strategy.

\begin{center}
\begin{figure}
\centerline{\includegraphics[width=12cm]{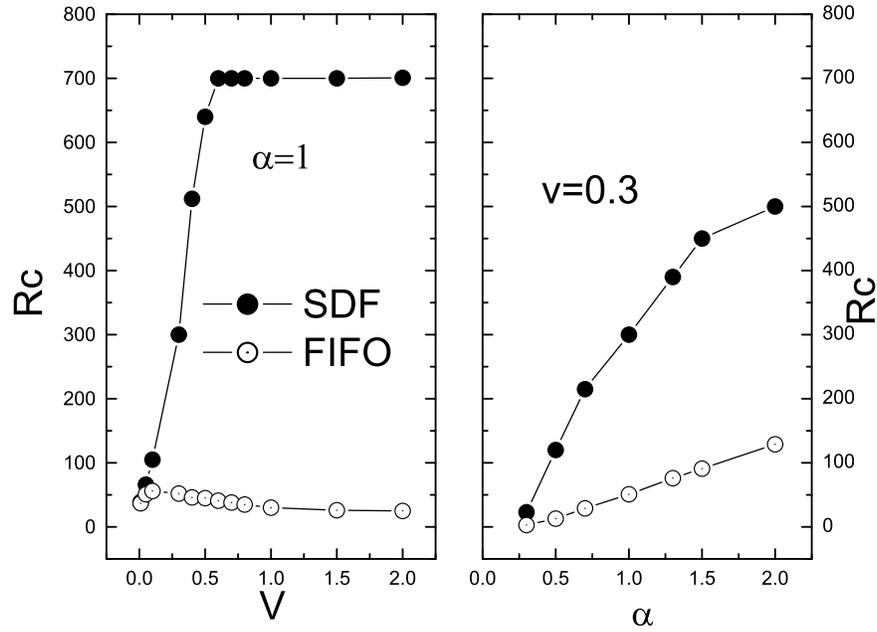}}
\vskip 0.2\baselineskip
\begin{flushleft}
\caption{ (a) The critical value $R_c$ as a function of the moving speed $v$ for the communication radius $\alpha=1$.  (b) The critical value $R_c$ as a function of $\alpha$ for moving speed $v=0.3$. Here $N=800$, $L=10$, $C=1$.}
\end{flushleft}
\end{figure}
\end{center}

Figure 3 represents the dependence of average end-to-end delay $<D>$ on $R$, and the insert shows that in the free-flow state($R<R_c$). We find $<D>$ of two strategies increases as $R$ increases. From the insert, we can see  $<D>$ of the shortest-distance-first strategy is a little lower than that of the first-in-first-out strategy in the free-flow state. However, form Fig. 3, $<D>$ of the shortest-distance-first strategy is much lower than that of the first-in-first-out strategy  in the congested state.

\begin{center}
\begin{figure}
\centerline{\includegraphics[width=12cm]{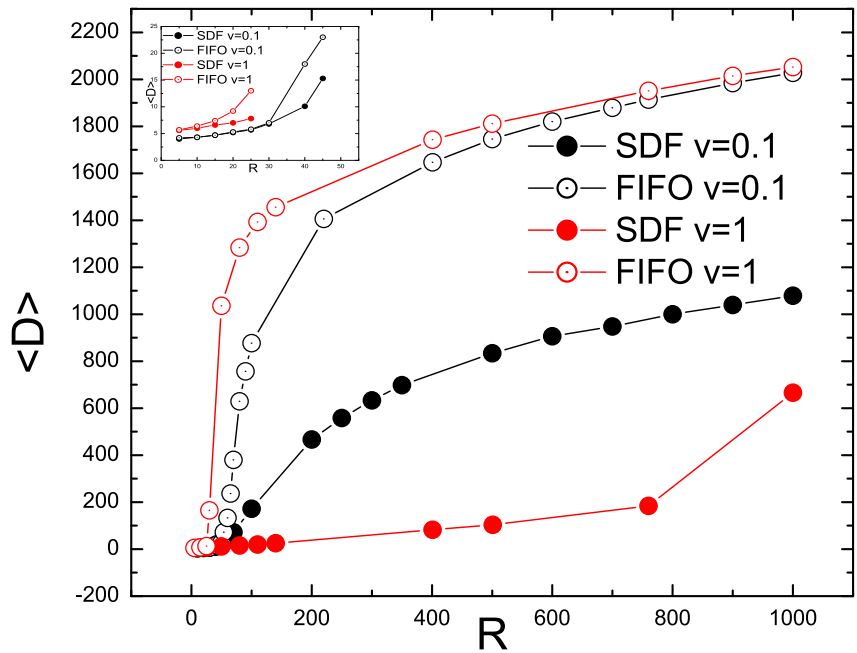}}
\vskip 0.2\baselineskip
\begin{flushleft}
\caption{The average end-to-end delay $<D>$  as a function of the packet-generation rate R. The inset shows $<D>$ as a function of R when $R<R_c$. Here $N=800$, $L=10$, $\alpha=1$, and $C=1$.}
\end{flushleft}

\end{figure}
\end{center}

Figure 4(a) and 4(b) show the dependence of the rate of queueing delay $Q$ and the packet arriving rate $A$ on $R$, respectively. From Fig. 4(a), $Q$ of two strategies is close to 0 when $R$ is small, and with $R$ increase, $Q$ increases quickly, when R increases to $\sim80$, $Q$ of the first-in-first-out strategy is close to the maximal value $1$, while that of the shortest-distance-first strategy is just $\sim 0.46$. From Fig. 4(b), in the free-flow state, the packet arriving rate $A$ of the two strategies is close to 1, while in the congestion state, $A$ of the first-in-first-out strategy decreases more quickly than that of the shortest-distance-first strategy, when $R$ increases to $1000$, $A$ of the first-in-first-out strategy is close to the minimal value $0$ while that of the shortest-distance-first strategy is $\sim 0.53$ and $\sim 0.8$ at $v=0.1$ and $1$, respectively. It is obvious that the shortest-distance-first strategy can achieve higher value of packet arriving rate and get lower value of the rate of queueing delay than the first-in-first-out strategy.

\begin{center}
\begin{figure}
\centerline{\includegraphics[width=12cm]{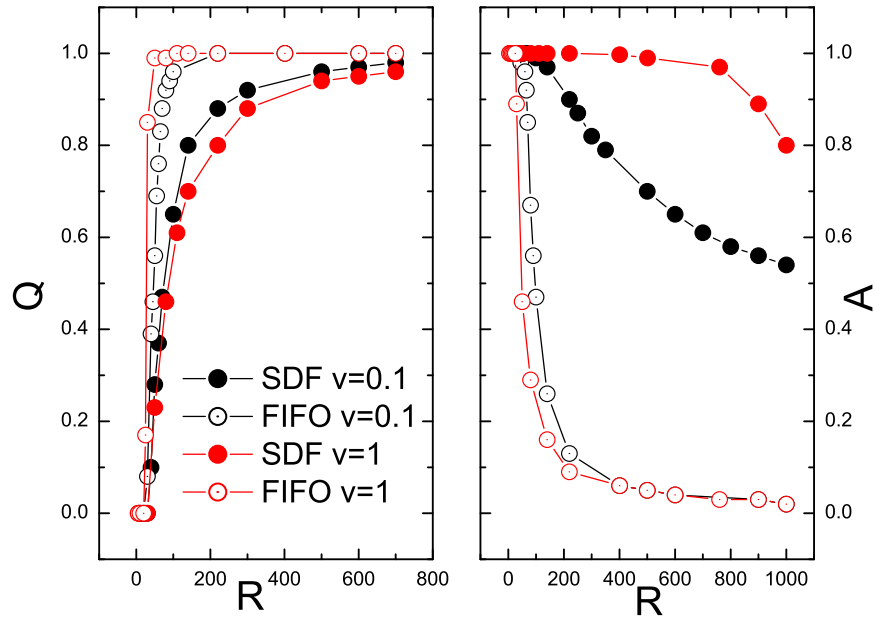}}
\vskip 0.2\baselineskip
\begin{flushleft}
\caption{ (a) The rate of queueing delay $Q$ as a function of the packet-generation rate R. (b) The packet arriving rate $A$ as a function of the packet-generation rate R. Here $N=800$, $L=10$, $\alpha=1$, and $C=1$.}
\end{flushleft}

\end{figure}
\end{center}

\section{Conclusions}
Traffic on mobile networks is a challenging work. However, previous routing strategies have little attention what order should the packets be forwarded, just simply used first-in-first-out queue discipline.
Based on queueing theory, we propose a shortest-distance-first strategy, in which packets those have shortest distance are delivered first, regardless of their arrival time.

Our strategy improves remarkably network throughput than the first-in-first-out strategy, especially when agents move at high speed. In addition, our strategy increases the packet arriving rate, and reduces average end-to-end delay and the rate of queueing delay.

We also find critical packet generating rate $R_c$ of the  first-in-first-out strategy increases as moving speed increases at the beginning and reaches the maximin value at $v=0.1$ and then decreases slowly, while that of the shortest-distance-first strategy increases until $v=0.7$ and then remains stable ($\sim700$). Besides, $R_c$ of the two strategies increases as communication radius increases. Finally, it should be pointed out the priority queue discipline can be applied to other routing strategies. Our work may be helpful in routing strategy designing on mobile networks.

\begin{center}
\textbf{Acknowledgements}
\end{center}
The work is supported by the National Science Foundation of China under Grant Nos 10975099, the Program for Professor of Special Appointment (Orientational Scholar) at Shanghai Institutions of Higher Learning under Grant Nos D-USST02, and the Shanghai project for construction of discipline peaks under Grant No.USST-02-SAI.

\end{document}